\documentclass[twoside,a4paper,11pt]{sca}
\usepackage{graphicx}
\usepackage{hyperref}
\usepackage{movie15}
\usepackage{natbib}  
\begin{document}
\pagenumbering{arabic}
\pagestyle{myheadings}
\thispagestyle{empty}
{\flushright\includegraphics[width=\textwidth,bb=90 650 520 700]{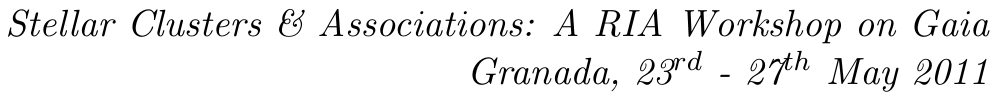}}
\vspace*{0.2cm}
\begin{flushleft}
{\bf {\LARGE
{\em Gaia} and $\sigma$~Orionis from 20\,$M_\odot$ to 3\,$M_{\rm Jup}$: \\
the most complete and precise Initial Mass Function with a parallax determination?
}\\
\vspace*{1cm}
J. A. Caballero
%
}\\
\vspace*{0.5cm}
Centro de Astrobiolog\'ia (CSIC-INTA), PO Box 78, E-28691 Villanueva de la Ca\~nada, Madrid, Spain
\end{flushleft}
\markboth{
Initial Mass Function, $\sigma$~Orionis and {\em Gaia}
}{ 
J. A. Caballero
}
\thispagestyle{empty}
\vspace*{0.4cm}
\begin{minipage}[l]{0.09\textwidth}
\ 
\end{minipage}
\begin{minipage}[r]{0.9\textwidth}
\vspace{1cm}
\section*{Abstract}{\small
The $\sigma$~Orionis cluster is to date the star-forming region with the largest number of confirmed brown dwarfs and substellar objects below the deuterium burning mass limit. 
The most massive star, $\sigma$~Ori~Aa, just in the cluster centre, is the $\sim$20\,$M_\odot$-mass O9.5V star that illuminates the Horsehead Nebula, while the least massive object yet reported, S\,Ori\,70, is only around 3\,$M_{\rm Jup}$. 
In the middle, there is a continuum of stars and substellar objects of all types (including magnetically active B2Vp stars, Herbig-Haro objects, FU\,Ori stars or T\,Tauri brown dwarfs) that makes the cluster a cornerstone in the study of the initial mass function, disc presence, X-ray emission or accretion at all mass domains.
However, the derived masses strongly depend on the actual heliocentric distance to the cluster. 
{\em Gaia} will solve the dilemma.
\normalsize}
\end{minipage}

\section{$\sigma$~Orionis: significance \label{section.significance}}

The {$\sigma$~Orionis} cluster ($\tau \sim$ 3\,Ma, $d \sim$ 385\,pc, $A_V <$ 0.3\,mag) is located in the easternmost part of the Ori\,OB1b association and is one of the most attractive and visited regions for night sky observers \citep{Ga67, Wo96, Wa08, Ca08c}.
The cluster gets the name from the homonymous massive star system in its centre, which is the fourth brightest star in the Orion Belt. 
The $\sigma$~Orionis cluster is important for several reasons:

\begin{figure}[ht]
\center
\includegraphics[width=0.99\textwidth]{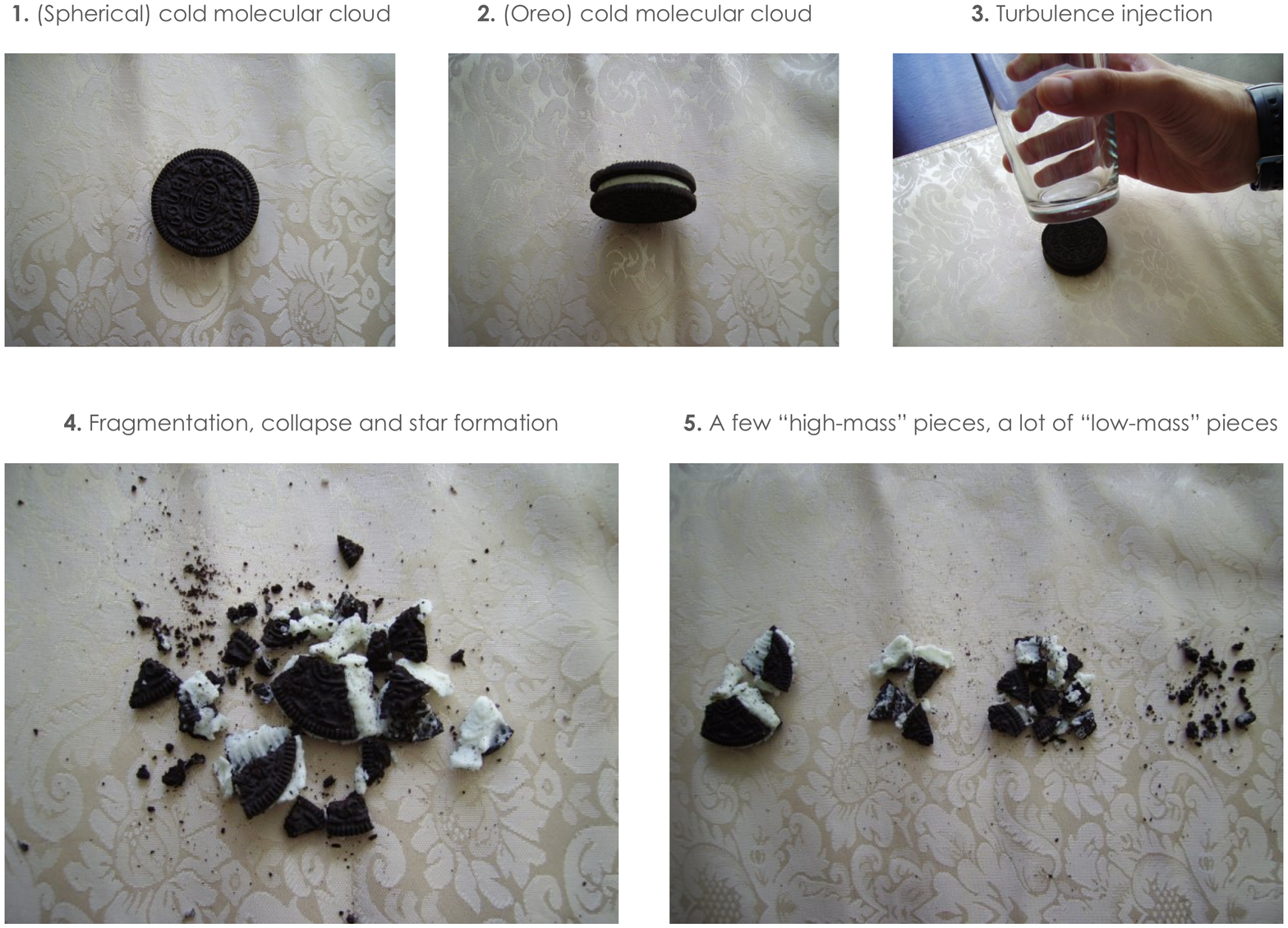} 
\caption{\label{fig.cookieIMF} Pictorical sketch of the star formation process and cluster IMF determination (the original idea of breaking into pieces, ``fragmenting'', a cookie and counting the pieces was first stated by V. J. S. B\'ejar; all pieces were eaten after the making of this figure).  
}
\end{figure}

\begin{itemize}
\item Its stars illuminate the {Horsehead Nebula} photodissociation region \citep{Po03, Pe05, Go06}.
\item It contains an abundant X-ray emitter population \citep{Fr06, Sk08, Ca+10}.
\item It is a cornerstone for studying disc frequency at $\tau \sim$ 3\,Ma at all mass domains \citep{Ol04, Ca+07, ZO07, Lu08}.
\item The helium-rich, magnetically active, B2Vp star {$\sigma$~Ori\,E} is in its centre \citep{Wa74, GH82, To05}.
\item It holds four Herbig-Haro objects and dozens Herbig Ae/Be, T~Tauri and FU~Ori stars  \citep{HM53, Re98, An04}.
\item The central star system is the most massive ``binary'' with an astrometric orbit \citep{Ha96, Ma98, Ca08b}.
\item Its proximity, youth and low extinction facilitate studies of accretion rates and frequency at low masses \citep{SE04, Ke05, Ga08}.
\item It is the star-forming region with the largest number of confirmed brown dwarfs ($M <$ 70\,$M_{\rm Jup}$) and objects below the deuterium burning mass limit (i.e., isolated planetary-mass objects; $M <$ 13\,$M_{\rm Jup}$) with spectroscopy and youth features, e.g., Li~{\sc i} absorption, H$\alpha$ and X-rays emission, infrared excess, radial velocity \citep{Be99, ZO00, ByN01, Ri11}. 
\end{itemize}

\section{$\sigma$~Orionis: previous IMF works  \label{section.previousIMF}}

\begin{table}[] 
	\center
	\caption{IMF works in the $\sigma$~Orionis cluster.} 
	\smallskip
	\center
\begin{minipage}{1.0\textwidth}
\center
	\begin{tabular}{ l c c l}
	\hline\hline 
Work				& Mass interval			& $\alpha$	& Colour		\\
				& [$M_\odot$]			& 			& in Fig.~\ref{fig.allIMF} \\  [0.5ex]   
	\hline 	
\citealt{Be01}		& 0.1--0.015			& +0.6		& Red		\\  	
\citealt{Tej02}		& 0.45--0.025			& +1.2		& Orange (dashed) \\  
\citealt{Sh04}		& 0.2--1.0				& +2.7		& Not drawn	\\   	
Caballero 2006, \citeyear{Ca10a}	
				& 3--0.6				& +1.4		& Yellow		\\   	
				& 0.4--0.006			& +0.4		& Yellow		\\ 
\citealt{GG06}		& 0.072--0.007			& +0.6		& Not drawn	\\   	
\citealt{Ca07}		& 18--1.5				& +2.0		& Green		\\   	
\citealt{Ca+07}		& 0.11--0.006			& +0.6		& Not drawn	\\   	
				& 0.072--0.006			& +0.4		& Light blue	\\ 
\citealt{Lo09}		& 0.49--0.010			& +0.5		& Blue (dashed) \\   	
\citealt{Bi09}		& 0.012--0.0035		& +0.0:		& Dark blue	\\   	
\citealt{Ca09}		& 20--1				& +2.3		& Magenta	\\   	
				& 0.3--0.035			& +0.3		& Magenta	\\  [0.5ex]   
	\hline
	\end{tabular} 
	\end{minipage}
	\label{table.IMF}
\end{table}

The abundance of known substellar objects has led $\sigma$~Orionis to become a key region all over the sky to study the initial mass function (IMF), especially at {\em very} low masses.
The IMF is the empirical function that describes the mass distribution (the histogram of stellar masses) of a population of recently-born stars, and its determination in a cluster is illustrated in Fig.~\ref{fig.cookieIMF}.

\begin{figure}[ht]
\center
\includegraphics[width=0.99\textwidth]{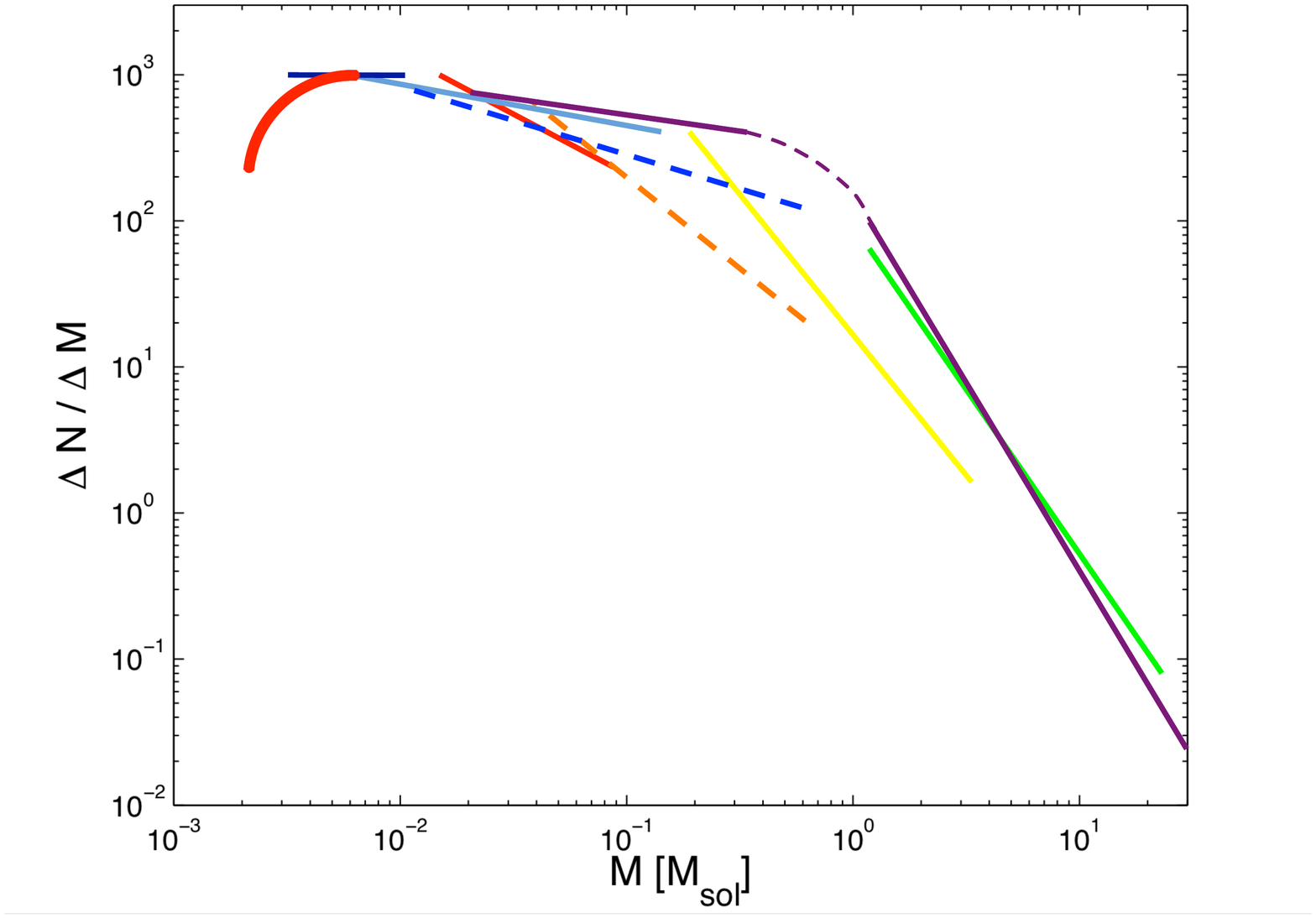} 
\caption{\label{fig.allIMF} Some mass spectra reported in the $\sigma$~Orionis cluster (see Table~\ref{table.IMF} for the references and colour code). 
}
\end{figure}

The behaviour of the IMF has been described repeatedly (e.g., \citealt{Kr02}), but $\sigma$~Orionis is probably the cluster where the IMF studies have gone lowest in mass with completeness and confidence, perhaps in more detail than in the Pleiades or the Orion Nebula Cluster. 
In Table~\ref{table.IMF}, I summarise all IMF works on $\sigma$~Orionis.
I tabulate the mass interval and the corresponding $\alpha$ index in the mass spectrum, $\Delta N / \Delta N = A M^{-\alpha}$ ($\alpha \equiv -\gamma \equiv 1-\Gamma$).
The general trend indicates a Salpeter's-like slope between 20 and 1\,$M_\odot$ ($\alpha \sim$ +2.35), a rather flat mass spectrum ($\alpha \sim$ +0.3) between 0.3 and 0.006\,$M_\odot$ (6\,$M_{\rm Jup}$) and a soft elbow in between, at 1--0.3\,$M_\odot$.
The mass spectrum may go {\em down} at a few Jupiter masses after the results exposed by \citet{PR11}.
See Fig.~\ref{fig.allIMF}.

\section{$\sigma$~Orionis: what to do next? \label{section.future}}

There are some caveats in the IMF determination in $\sigma$~Orionis:

\begin{itemize}

\item The IMF relies on cluster parameters and on the used theoretical models.
But which models to use? 
How to handle uncertainties at young ages?
Which band or combination of bands to use?
Are colour-dependent bolometric corrections better than apparent magnitudes only?
How to match different isochrones of different authors valid at different mass and effective temperature intervals?

\item How old is $\sigma$~Orionis?
There have been reports of ages from 1 to 10\,Ma, with a general consensus towards the 2--4\,Ma interval, being 3\,Ma the canonical value.
But what if it is slightly older or younger?
Is there any age spread? 
How do we account for the different accretion history?

\item What is the contamination rate in the cluster?
In $\sigma$~Orionis, there is contamination by B stars in nearby star-forming regions in Ori~OB1b, foreground low proper-motion AFG dwarfs, background giants, field late-M-, L- and T-type stars and brown dwarfs, red quasars...
Spectroscopy, astrometry and methane imaging help disentangling actual cluster members from interlopers \citep{Sa08, Ca+08, Ca10b, PR11}.

\item What is the ``system IMF''?
Further binarity studies at all mass and separation domains are required.
Some surprises await us, such as the confirmation that the $\sigma$~Ori star itself is not only an astrometric binary, but also a {\em massive spectroscopic triple} containing an O9.5V, a B0.5V and and early B dwarf (Sim\'on-D\'iaz et al., this volume).

\item Is there a cut-off of the IMF below 0.006\,$M_\odot$?
To ascertaint it, there is no other way that performing ultradeep imaging ($I >$ 26\,mag, $J >$ 22\,mag, $H >$ 22\,mag) in an area wider than 1000\,arcmin$^2$ with methane, astrometric and spectroscopic follow-up of T-type candidates, which requires a strong observational effort.

\item What is the distance to $\sigma$~Orionis?
Is it 350\,pc ({\em Hipparcos}, with large error bars), 450\,pc \citep{Sh08}, or something in between at around 400\,pc \citep{Ca08b, MN08}?
Solution: {\em Gaia}!

\end{itemize}


{\em Gaia} will likely not detect cluster brown dwarfs ($J  >$ 14.5\,mag), and some cluster dynamical studies (radial velocities, proper motions) will arrive earlier than the ESA mission, but {\em Gaia} will measure the parameter with the largest error contribution to the IMF, 
{\em distance}, from accurate parallax determination.
A list of dozens of confirmed cluster members with $V <$ 20\,mag is easy to accomplish; 
we only need to wait a couple of years.


%
\small  
\section*{Acknowledgments}   
This work was supported by the Spanish Ministerio de Ciencia e Innovaci\'on and the Comunidad de Madrid under under grants AYA2008-00695, AYA2008-06423-C03-03 and S2009/ESP-1496.
\bibliographystyle{aa}
\bibliography{Caballero_JA_sg}

\end{document}